\documentclass[pra,aps,twocolumn,superscriptaddress,longbibliography]{revtex4-1}
\usepackage[colorlinks=true, citecolor=red, urlcolor=blue ]{hyperref}

\usepackage{}
\usepackage{float}
\usepackage{epsfig}
\usepackage{graphicx}
\usepackage{amsfonts,amssymb}
\usepackage{amsmath}
\usepackage{physics}
\usepackage{color}
\usepackage {times}

\definecolor{refcolor}{rgb}{1.0,0.0,0.0}

\begin{document}

\title{Local entanglement transfer to multiple pairs of spatially separated observers}

\author{Tanmoy Mondal}
\affiliation{Harish-Chandra Research Institute,  A CI of Homi Bhabha National
Institute, Chhatnag Road, Jhunsi, Prayagraj 211019, India}
\author{Kornikar Sen}
\affiliation{Harish-Chandra Research Institute,  A CI of Homi Bhabha National
Institute, Chhatnag Road, Jhunsi, Prayagraj 211019, India}
\author{Chirag Srivastava}
\affiliation{Harish-Chandra Research Institute,  A CI of Homi Bhabha National 
Institute, Chhatnag Road, Jhunsi, Prayagraj 211019, India}
\affiliation{Laboratoire d'Information Quantique, CP 225, Université libre de Bruxelles (ULB), Av. F. D. Roosevelt 50, 1050 Bruxelles, Belgium}
\author{Ujjwal Sen}
\affiliation{Harish-Chandra Research Institute,  A CI of Homi Bhabha National
Institute, Chhatnag Road, Jhunsi, Prayagraj 211019, India}

\begin{abstract}
Entanglement is an advantageous but at the same time a costly resource utilized in various quantum tasks. 
For an efficient usage and deployment of entanglement, we envisage the scenario where a pair of spatially separated observers, Charu and Debu, want to share entanglement without interacting with each other. 
As a way out, their systems can separately and locally interact with those of Alice and Bob, respectively, who already share an entangled state.
%
We ask 
if it is possible to transfer entanglement from the Alice-Bob pair to multiple Charu-Debu pairs,
where the Alice-Bob pair only possesses a limited amount of pre-shared entanglement.
We find 
joint unitaries, which when applied by Alice and one of the Charus, 
%
and by Bob and the corresponding Debu,
such that a nonzero amount of the entanglement shared between Alice and Bob can be sequentially transferred to an indefinite number of pairs of Charus and Debus. 
We discuss the amount of entanglement that can be transferred to a fixed number of pairs using these unitaries. Also, we determine to how many pairs a fixed amount of entanglement can be transferred. Moreover, by optimizing over all possible local unitaries, we analyze the maximum number of pairs to which entanglement can be transferred in such a way that each pair gets at least a fixed amount of entanglement.
\end{abstract}

\maketitle
\section{Introduction}
Entanglement, one of the several interesting aspects of shared quantum systems, was discovered by Einstein, Podolsky, and Rosen~\cite{EPR1}.
It is a quantum correlation existing between two entities or between two degrees of freedom of the same entity~\cite{ent1,ent2}. This peculiar correlation, after its discovery, gained tremendous attention due, in part, to its numerous applications in quantum technologies, especially in quantum communication. Some examples of its applications are quantum teleportation~\cite{Teleportation}, quantum dense coding~\cite{Densed}, entanglement-based quantum cryptography~\cite{crypto1, crypto2},  entanglement swapping~\cite{swap1,swap2}, and randomness amplification~\cite{randomness}.
Thus, creation, detection,  distribution, and preservation of entanglement is, along with being a fundamental field of research, is also of crucial practical significance in quantum technologies.

The traditional method of creating entanglement between two parties is to bring them to the same site and apply a global operation on the complete system. But there can be situations where two distant parties, say Charu and Debu, urgently need shared entanglement to, say, teleport a quantum message or distribute secret keys, but do not have any entanglement shared between them. This problem of Charu and Debu can indeed be solved if another pair of observers, Alice-Bob, already share an entangled system among them and Charu (Debu) has access to Alice's (Bob's) part of the entangled system. By considering such situations, rather than resorting to the costly approach of physically bringing pairs of parties together, we want to analyze the extent to which it is possible to create entanglement between spatially separated observers using simple means of local operations and a shared pre-existing but limited amount of entanglement. The pair of observers, Charu and Debu, can locally operate on their individual systems and subsystems of Alice and Bob, respectively, to establish entanglement among their systems. This comes at the cost of at least a partial loss in the entanglement content in the Alice-Bob shared system. 
In this paper, we study a sequential entanglement transfer protocol where the main interest is in the number of Charu-Debu pairs who can establish a certain amount of entanglement - required to achieve their tasks - by locally interacting with the Alice-Bob pair, who initially shares a given amount of entanglement. We assume that the different pairs of Charu and Debu must extract entanglement from the Alice-Bob pair at different times, and thus they act sequentially. Further, 
we assume that different Charu-Debu pairs do not interact among themselves. An important question within this arena is whether an unbounded number of Charu-Debu pairs can transfer entanglement to their systems using the above protocol. 
The answer to this question is 
troubled by the fact that
each time any of the Charu-Debu pairs interacts with the  Alice-Bob pair, 
a loss of entanglement shared by Alice-Bob may occur, and at the same time, the interaction may establish entanglement between three more pairs of observers, viz., Charu-Debu, Alice-Charu, and Bob-Debu.  And, the entanglement established between Alice-Charu and Bob-Debu is useless for creating any entanglement between a subsequent Charu-Debu pair, as the different Charu-Debu pairs are assumed to act independently with the Alice-Bob pair. 
 


Recently, there has been a stream of works analyzing sequential tasks, such as sequential detection of resources like Bell nonlocality \cite{Silva15,Mal16,Saha19,Das19,Brown20,cabello20,Fei21} and entanglement \cite{Bera18,sriv21,Srivastava22,Pandit22,srivastava_GME1_22,srivastava_GME2_22}. These protocols have also been extended to quantum networks \cite{Mao22}. They have found applications in device-independent randomness generation~\cite{Curchod17}, and are also useful as state preparations can in principle be more costly than utilizing a given system multiple times.
There also exists a different - and earlier - perspective of manipulating and transferring entanglement among observers, known as the splitting of entanglement~\cite{ent-splitting}. Specifically, a situation is considered where two parties, say Alice and Bob, share an entangled state, and another observer Charu, applies joint unitary operations on Bob's and his systems and splits this entanglement into a branch. Hence, the final three-party state shared between Alice, Bob, and Charu consists of entanglement between Bob and Alice as well as Bob and Charu. This has then been extended to splitting into more than one branch.


In this paper, we ask 
if it is possible to transfer some of the entanglement of the Alice-Bob pair to a new pair of parties, say Charu$^{(1)}$ and Debu$^{(1)}$ by considering a situation where Alice and Bob initially share a maximally entangled state of two qubits. We observe that the transmission is indeed possible with the operation of joint unitaries on Alice's and Charu$^{(1)}$'s subsystems and on Bob's and Debu$^{(1)}$'s subsystems. Therefore, 
we can successfully create entanglement between Charu$^{(1)}$ and Debu$^{(1)}$ without relying on any interaction,
be it classical or quantum, between Charu$^{(1)}$ and Debu$^{(1)}$. The crucial resource used here is the entanglement between Alice and Bob. The amount of transmission - of entanglement from Alice-Bob to Charu$^{(1)}$-Debu$^{(1)}$ - of course, depends on the choice of the joint unitaries. 
We find that there always exists a pair of joint unitaries corresponding to every chosen resultant amount of entanglement shared between Charu$^{(1)}$ and Debu$^{(1)}$.
Moreover, we realize that after a partial transmission of entanglement to Charu$^{(1)}$ and Debu$^{(1)}$, the remainder of entanglement shared between Alice and Bob can again be transferred - again possibly partially - to another pair, say Charu$^{(2)}$ and Debu$^{(2)}$. 
We then obtain the maximum number of pairs that can receive entanglement while ensuring each pair obtains a fixed minimum amount of entanglement by optimizing over local joint unitaries. We observed that, in principle, the sequence of transmission of entanglement to different pairs can be repeated an unbounded number of times using some specific unitaries. By fixing such a specific form of unitary, we first obtain the maximum amount of entanglement that can be extracted by each Charu-Debu pair, given that the number of such pairs is fixed and each pair extracts an equal amount of entanglement. Then, we also determine the maximum number of pairs that can extract a fixed amount of entanglement.

%
%
%

The rest of the paper is organized as follows. In Sec. \ref{sec_2}, we elucidate the entanglement transfer protocol from a maximally entangled source to $n$ pairs of 
separable
states, providing an estimation of the value of $n$ through optimization over the local joint unitaries. To delve deeper into this phenomenon, we adopt a specific form of unitary and concentrate on entanglement transfer to the initial pair in Sec. \ref{sec3}. For the chosen unitary form, Sec. \ref{sec_4} presents a numerical investigation of entanglement transfer to several pairs. Precisely, we determine to how many pairs a fixed amount of entanglement can be transferred, and if we fix the number of pairs to which entanglement is required to be shifted, then what is the maximum amount of entanglement that each of the pairs would receive? Subsequently, in Sec. \ref{sec5}, we derive the analytical conditions for sequential entanglement transfer to an arbitrarily large number of separable pairs and speculate on the feasibility of such a process.
We present a conclusion in Sec.~\ref{harkin-choT}.

\section{Entanglement Transfer Protocol} \label{sec_2}
Let Alice and Bob initially share a two-qubit maximally entangled state $\rho_{AB}^{(0)}$, given by $\rho_{AB}^{(0)}=\ket{\phi^+}\bra{\phi^+}$, where $\ket{\phi^+}$ has the form $\frac{1}{\sqrt{2}}(\ket{00}+\ket{11})$. There are $n$ pairs of distant parties, say Charu$^{(n)}$ and Debu$^{(n)}$, where Charu$^{(n)}$ can easily access Alice's lab and similarly, Debu$^{(n)}$ can easily enter Bob's lab. The index $(n)$ represents the $n-$th pair of Charu and Debu. Let us assume the initial states of each Charu$^{(n)}$ and Debu$^{(n)}$ are $\rho_C^{(n)}=\ket{0}\bra{0}$ and $\rho_D^{(n)}=\ket{0}\bra{0}$. Thus the joint state of Charu$^{(n)}$ and Debu$^{(n)}$ is $\rho_{CD}^{(n)}=\ket{0}\bra{0}\otimes\ket{0}\bra{0}$. The Hilbert spaces on which Alice, Bob, Charu$^{(n)}$, and Debu$^{(n)}$'s states act are respectively denoted by  $\mathcal{H}_{A}$, $\mathcal{H}_B$, $\mathcal{H}_C^{(n)}$, and $\mathcal{H}_D^{(n)}$, all of which have same dimension $d$. 

Let us now consider that the first pair, i.e., Charu$^{(1)}$ and Debu$^{(1)}$ interacts with Alice and Bob respectively. The unitaries representing the evolution of the joint state, $\rho_C^{(1)}\otimes\rho_{AB}^{(0)}\otimes\rho_D^{(1)}$, under the interaction, are $U_{CA}^{(1)}$ and $U_{BD}^{(1)}$. Hence the final state after the evolution can be written as 
\begin{equation}
\rho'^{(1)}_{CABD}=U_{CABD}^{(1)}\left(\rho_C^{(1)}\otimes\rho_{AB}^{(0)}\otimes\rho_D^{(1)}\right)\left(U_{CABD}^{(1)}\right)^\dagger, \label{eq6}
\end{equation}
where $U_{CABD}^{(1)}=\left(U_{CA}^{(1)}\otimes U_{BD}^{(1)}\right)$. We want to examine if the final state shared between Charu$^{(1)}$ and Debu$^{(1)}$, i.e., $\rho^{(1)}_{CD}=\Tr_{AB}\left({\rho'^{(1)}_{CABD}}\right)$ contains non-zero amount of entanglement, say $E_{CD}^{(1)}$. 

Because of the interaction, the bipartite state of Alice and Bob's system has been changed to $\rho^{(1)}_{AB}=\Tr_{CD}\left({\rho'^{(1)}_{CABD}}\right)$. But there can be if not all, some entanglement left between Alice and Bob's shared state, $\rho^{(1)}_{AB}$. Let us say the amount of entanglement is $E_{AB}^{(1)}$. In such a situation, a new pair comes, say Charu$^{(2)}$ and Debu$^{(2)}$, and interacts in the same way with Alice and Bob. The corresponding unitaries which demonstrate the interaction are $U_{CABD}^{(2)}=\left(U_{CA}^{(2)}\otimes U_{BD}^{(2)}\right)$. The protocol can be repeated multiple times in this way. The composite state of the four system shared between Charu$^{(n)}$, Alice, Bob and Debu$^{(n)}$, exactly after the interaction of Alice and Bob with Charu$^{(n)}$ and Debu$^{(n)}$, respectively, can be written as 
\begin{equation}
\rho'^{(n)}_{CABD}=U_{CABD}^{(n)}\left(\rho_C^{(n)}\otimes\rho_{AB}^{(n-1)}\otimes\rho_D^{(n)}\right)\left(U_{CABD}^{(n)}\right)^\dagger.\label{eq3}
\end{equation}
The unitary $U_{CABD}^{(n)}$ acts locally on the systems, i.e., $U_{CABD}^{(n)}=U_{CA}^{(n)}\otimes U_{BD}^{(n)}$. Initially Charu$^{(n)}$ and Debu$^{(n)}$ were separated for all $n$. Our goal is to investigate whether the aforementioned process can transfer a non-zero portion of the entanglement from $\rho^{(n-1)}_{AB}$ to Charu$^{(n)}$ and Debu$^{(n)}$, and if so, to determine how the amount of entanglement in each pair, denoted as $E_{CD}^{(n)}$ varies with its corresponding number, $n$.

Throughout the paper, we constrain ourselves with the condition that $U_{CA}^{(n)}=U_{BD}^{(n)}=U_{CA}^{(1)}=U_{BD}^{(1)}=U$ for all $n$. In order to investigate the potential for transferring entanglement, we determine, by optimizing over the unitaries, $U$, the maximum number of pairs, i.e., $n$, to whom Alice and Bob can transfer a particular amount of entanglement which is bounded below by a value, say $2^{-x}$, where $x$ is any fixed positive real number. In particular, by using a non-linear optimization algorithm, NLopt, we maximize $n$ by constraining $E_{CD}^{(n)}>2^{-x}$. We restrict ourselves to qubit systems and use logarithmic negativity to measure the entanglement contained in the system shared between Charu$^{(n)}$ and Debu$^{(n)}$, $E_{CD}^{(n)}$.

\begin{figure}[h]
\centering
\includegraphics[scale=0.40]{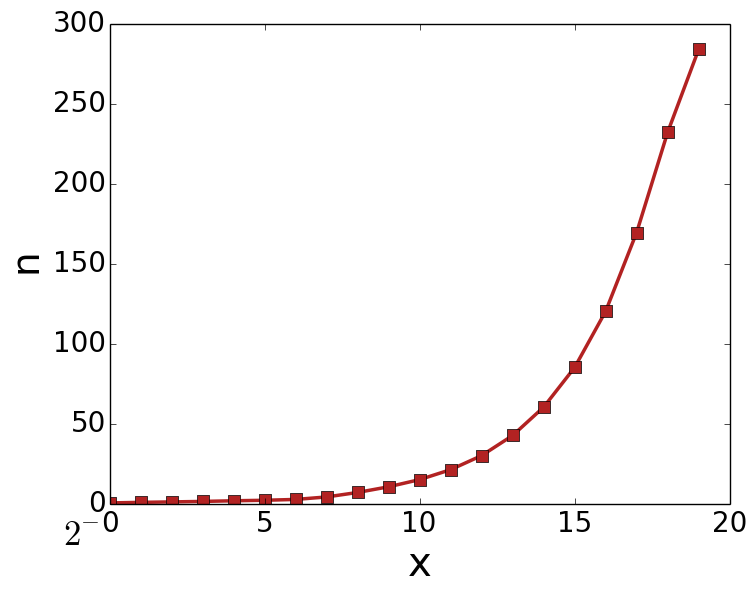}
	\caption{Successive transformation of entanglement to various pairs. In the vertical axis we plot the maximum number of pairs to whom Alice and Bob can hand over entanglement so that the parties receiving entanglement can have at least $2^{-x}$ amount of entanglement whereas, in the horizontal axis, we plot the corresponding value of $x$. The vertical axis is dimensionless and the horizontal axis has units of $\log ebits$. We plot the corresponding data points using black squares. We have connected the black squares with black lines to track the behavior.}
	\label{fig_opt}
\end{figure}

In Fig. \ref{fig_opt}, we plot the optimal value of $n$ with respect to $x$. The figure unambiguously demonstrates that it is feasible to transfer a minute but still a non-zero finite quantity of entanglement to an arbitrarily large number of pairs. This behavior fascinates us to look into the matter in a comprehensive way. We will present a formal proof of the statement in Sec. \ref{sec5}. But before going into that, in order to observe the behavior of entanglement transportation in more detail, we first focus on the entanglement transfer to the first pair only.

\section{A Close Look at Entanglement Transfer to a  Single Pair}
\label{sec3}
To investigate further, we focus on a particular form of unitary, for the entangle transfer. A unitary can, in general, be represented as $U= \exp (-iH t)$, where $H$ denotes a Hamiltonian. Without going into the complicacy of the Hamiltonian, we chose a simple spin Hamiltonian of the form 
\begin{eqnarray}
    {H}=\gamma(\sigma _x \otimes \sigma _x +\sigma _y \otimes \sigma _y), \label{eq5}
\end{eqnarray}
where $\gamma$ is the coupling constant representing the strength of the interaction between two qubits. By considering $U^{(1)}_{CA}=U^{(1)}_{BD}=\exp (-iHt)$ we analyze the behavior of transferred entanglement, $E_{CD}^{(1)}$, between Charu$^{(1)}$ and Debu$^{(1)}$ given by 
\begin{eqnarray}
    E^{(1)}_{CD}=\log\left[\frac{1}{8}\left(11-4\cos(4\lambda)+\cos(8\lambda)\right)\right],\label{eq2}
\end{eqnarray}
where we introduce a unit less parameter $\lambda=\gamma t$, $t$ indicating the total time of interaction. We name $\lambda$ as ``interaction strength". We also calculate the remaining entanglement between Alice and Bob, $E^{(1)}_{AB}$, as a function of $\lambda$.

\begin{figure}[h]
    \centering
    \includegraphics[scale=0.4]{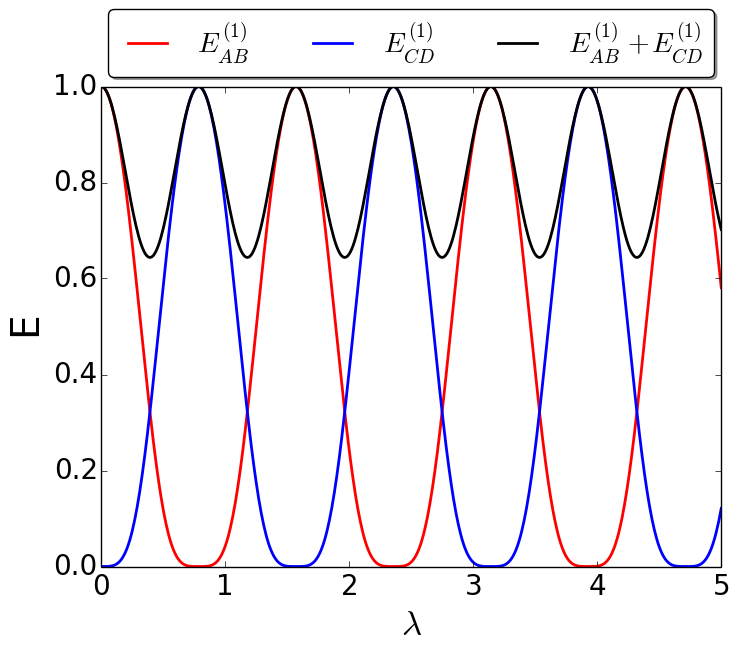}
    \caption{Behaviour of shared entanglement with respect to interaction strength. The entanglement contained in the systems shared between Alice and Bob is plotted in red, between Charu$^{(1)}$ and Debu$^{(1)}$ is in blue, along the vertical axis. In contrast, the horizontal axis represents the corresponding value of $\lambda$. The black curve exhibits the behavior of the sum of the above two entanglements with $\lambda$. The horizontal axis is dimensionless while the vertical axis has the dimension of e-bits.}
    \label{fig_1_pair_ent}
\end{figure}
We depict the nature of $E_{CD}^{(1)}$, $E_{AB}^{(1)}$, and $E_{AB}^{(1)}+E_{CD}^{(1)}$ in Fig. \ref{fig_1_pair_ent}. The blue line in the figure, representing the amount of entanglement Charu$^{(1)}$ and Debu$^{(1)}$ received, shows a periodic behavior with interaction strength, $\lambda$, due to the periodic nature of the unitary. The remaining entanglement between Alice and Bob, $E_{AB}^{(1)}$, is plotted in red. It shows a complimentary behavior with the same periodicity. At $t=0$, $\rho_{AB}^{(1)}$ is maximally entangled showing $E_{AB}^{(1)}=1$ in the figure and $\rho_C^{(1)}$ and $\rho_D^{(1)}$ begin separated, thus $E_{CD}^{(1)}=0$. Initially, $E^{(1)}_{AB}$ decreases, and $E_{CD}$ increases up to a certain value of $\lambda$, say $\lambda_c$, at which $E^{(1)}_{AB}$ becomes equal to $0$ and $E^{(1)}_{CD}$ reaches maximum value, $1$. It indicates that the initial entanglement of $\rho^{(0)}_{AB}$ has completely been transferred to Charu$^{(1)}$ and Debu$^{(1)}$ making them maximally entangled. As a result, the systems of Alice and Bob become separable. If the value of $\lambda$ is increased more, the process gets reversed, i.e., Charu$^{(1)}$ and Debu$^{(1)}$'s state starts to return the entanglement back to Alice and Bob, until the initial condition is reached. The process keeps on repeating in this way with $\lambda$. The sum of the entanglement of two pairs of parties, i.e., $E^{(1)}_{AB}+E^{(1)}_{CD}$, shown using the green line in Fig. \ref{fig_1_pair_ent}, also has the oscillatory behavior but its value always remains less than a unit which is obvious because local operations can not create or increase overall entanglement. We can see some loss of entanglement in the process for most of the $\lambda$ values which actually gets transferred between Alice and Charu$^{(1)}$ and/or Bob and Debu$^{(1)}$. \\

\section{Sequential Transfer of Entanglement to Multiple Pairs} \label{sec_4}
Let us again focus on the scenario involving multiple pairs of parties. In particular, Alice and Bob still share an entangled state and want to transfer their entanglement to others. But instead of being restricted to one pair of receivers, they want to transfer their entanglement to multiple pairs of parties in sequential order. 
We want to examine the amount of received entanglement by the other successive pairs. 

In this aim, let us consider the same initial states, that is, $\rho^{(0)}_{AB}=\ket{\phi^+}\bra{\phi^+}$, $\rho_C^{(1)}=\ket{0}\bra{0}$, $\rho_D^{(1)}=\ket{0}\bra{0}$. Instead of optimizing overall unitaries, let us consider the same pair of unitaries, i.e., $U_{AC}^{(n)}=U_{BD}^{(n)}=\exp(-i\lambda(\sigma _x \otimes \sigma _x +\sigma _y \otimes \sigma _y))$ for all $n$. After the application of these unitaries, Alice and Bob's shared state becomes $\rho_{AB}^{(1)}=$
\begin{eqnarray*}
\left(\begin{matrix}
\frac{1}{2}(1+\sin^4(2\lambda))&0&0&\frac{1}{2}\cos^2(2\lambda)\\
0&\frac{1}{8}\sin^2(4\lambda)&0&0\\
0&0&\frac{1}{8}\sin^2(4\lambda)&0\\
\frac{1}{2}\cos^2(2\lambda)&0&0&\frac{1}{2}\cos^4(2\lambda)
\end{matrix}\right)
\end{eqnarray*}
The corresponding amount of entanglement acquired by Charu$^{(1)}$, and Debu$^{(1)}$ is expressed in Eq \eqref{eq2}. 
Now, Charu$^{(2)}$ and Debu$^{(2)}$ comes with the states $\rho_C^{(2)}=\ket{0}\bra{0}$ and $\rho_D^{(2)}=\ket{0}\bra{0}$ acted upon by the unitaries, $U_{CA}^{(2)}$ and $U_{BD}^{(2)}$, receives entanglement of the following amount, 
\begin{equation}
    E_{CD}^{(2)}=\log_2[(131-4\cos (8\lambda)+\cos(16\lambda))/128].\nonumber
\end{equation}

\begin{figure}
\centering
\includegraphics[scale=0.40]{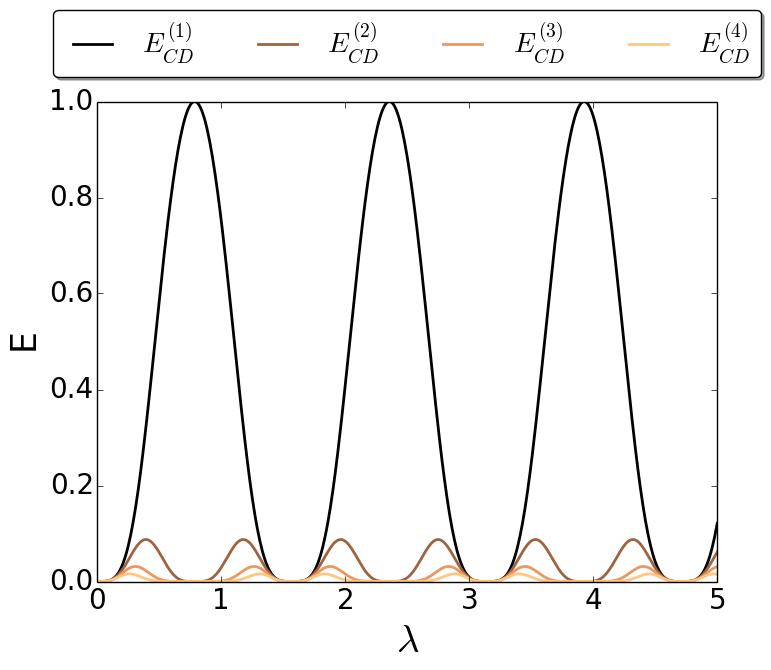}	\caption{Transmitting entanglement to multiple pair of systems. In the vertical axis we plot the amount of entanglement transferred to the state shared between Charu$^{(n)}$ and Debu$^{(n)}$ for $n=$1 (black line), 2 (brown line), 3 (orange line), and 4 (yellow line) as a function of the coupling $\lambda$ which is represented in the horizontal axis. The vertical axis is in ebits whereas the horizontal axis is dimensionless. The mathematical form of the entanglement plotted using black and brown color are expressed in Eqs. \eqref{eq2} and \eqref{eq3} respectively. The mathematical form of the orange and yellow curve can be found by following the same method described in Sec. \ref{sec3}.}
	\label{fig2}
\end{figure}

In Fig. \ref{fig2}, we plot the entanglement transferred in this way between Charu$^{(n)}$ and Debu$^{(n)}$ for $n=$1, 2, 3, 4. It is noticeable that all the entanglements are oscillatory functions of $\lambda$ which is also evident from the mathematical expressions of $E_{CD}^{(1)}$ and $E_{CD}^{(2)}$ that are trigonometric functions of $\lambda$. From the figure, it is visible that a half-wavelength of $E_{CD}^{(1)}$ contains two half-wavelengths of $E_{CD}^{(2)}$, which is evident because when $E_{CD}^{(1)}$ is maximum, it absorbs the complete entanglement contained between Alice and Bob and thus there is no entanglement left within Alice and Bob to transfer to the next pair, i.e.,  Charu$^{(2)}$ and Debu$^{(2)}$. It can also be observed that after providing some entanglement to the first pair, 
i.e., whatever the amount of leftover entanglement, $E_{AB}^{(1)}$, the application of these fixed unitaries can not extract all of it and transfer it to any of the next pairs. That is, one needs to either change the parameter, $\lambda$ of unitary at every round or change the form of unitary itself, in order to extract the total remained entanglement of Alice-Bob.
\begin{figure}[h!]
\centering
\includegraphics[scale=0.40]{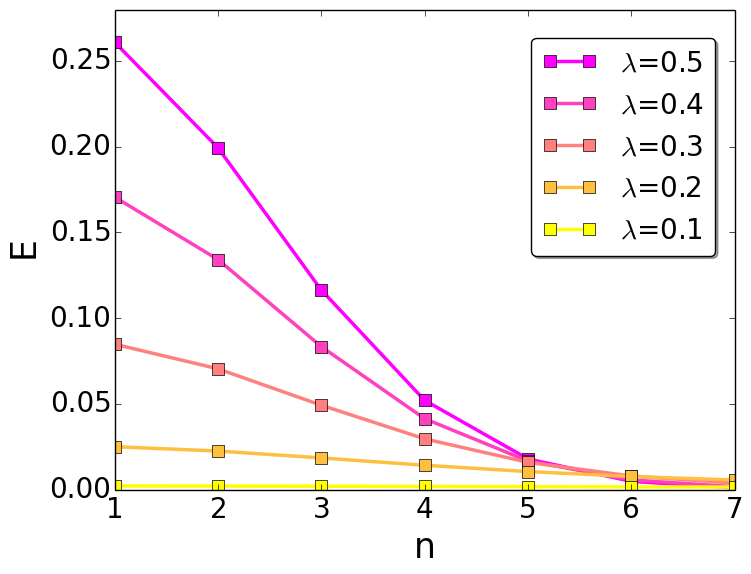}
	\caption{How many successive entanglement transfer is possible? Along the vertical axis, we plot the amount of entanglement that can be transferred between Charu$^{(n)}$ and Debu$^{(n)}$ with respect to $n$ where the corresponding value of $n$ is represented along the horizontal axis. The vertical axis is in ebits whereas the horizontal axis is dimensionless. The curves are corresponding to three different couplings, i.e., $\lambda$, as specified in the legend.}
	\label{fig3a}
\end{figure}

\begin{figure}[h!] \label{fig_n_vsx}
\centering
\includegraphics[scale=0.40]{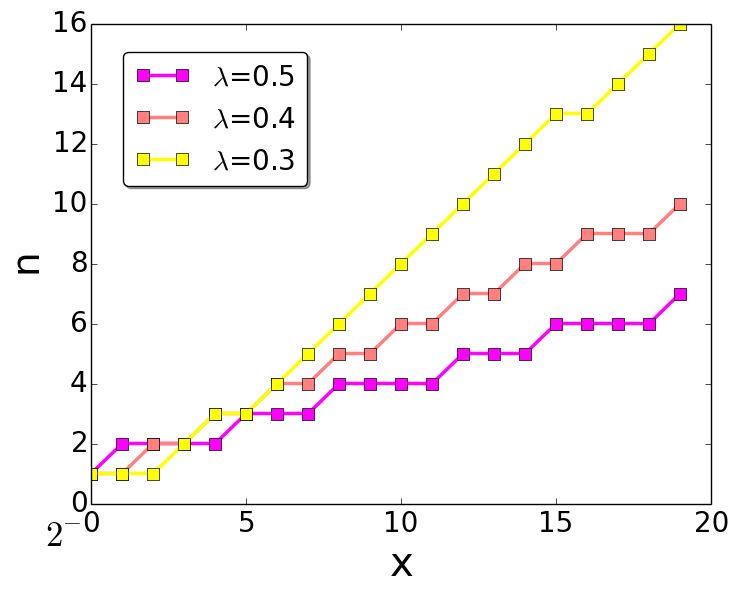}
	\caption{Supplying a minimum amount of entanglement to multiple pairs of parties. In the vertical axis we plot the maximum number of pairs to whom Alice and Bob can hand over entanglement so that the parties receiving entanglement can have at least $2^{-x}$ amount of entanglement whereas, in the horizontal axis, we plot the corresponding value of $x$. The vertical axis is dimensionless and the horizontal axis has units of $\log ebits$. The three curves are corresponding to three different couplings, i.e., $\lambda$, as specified in the legend.}
	\label{fig3}
\end{figure} 
Though the amount of entanglement has qualitatively similar behavior for $n=2$, 3, and 4, the quantitative behaviors are strictly distinct. To analyze the dependence of the amount of transferred entanglement on $n$, in Fig. \ref{fig3a}, we plot the amount of entanglement shared between Charu$^{(n)}$ and Debu$^{(n)}$, $E_{CD}^{(n)}$, after their interaction with Alice and Bob with respect to $n$ for different values of $\lambda$. From the plot, we see even for large enough values of $n$, a small but non-zero amount of entanglement is being transferred between Charu$^{(n)}$ and Debu$^{(n)}$. Next, we try to determine the number of pairs of Charu$^{(n)}$ and Debu$^{(n)}$, i.e., $n$, to whom Alice and Bob can transfer at least $2^{-x}$ amount of entanglement where $x$ is any positive real number. In this regard, in Fig. \ref{fig3}, we plot $n$ as a function of $x$. Up to a certain cut-off, $\lambda_c$, beyond which $E_{CD}^{(n)}$ starts to decrease again, the amount of entanglement transfer increases with the increase of the interaction strength, $\lambda$. So, by tuning $\lambda$ we can control the amount of entanglement that we want to transfer to Charu and Debu in each iteration. As a consequence, smaller values of $\lambda$ allow Alice and Bob to transfer entanglement to a larger number of pairs. This trend is evident from Fig. \ref{fig3} which exhibits a monotonic increase of $n$ with the increases with $x$. We discuss this behavior in the next section in more detail.

\section{An arbitrary number of pairs can extract entanglement from a source of fixed entanglement}
\label{sec5}
Motivated by the numerical results demonstrated in the previous section, we aim to explore by analytical means 
how large a number of pairs, having separable states initially, it is possible to provide any non-zero finite amount of entanglement. In this context we state the following theorem:
\vspace{2mm}\\
\textbf{Theorem 1.} \textit{Corresponging to every positive integer, $N$, there exists a positive real number, $E$, and at least one pair of unitaries such that each of $N$ pairs of Charu$^{(n)}$ and Debu$^{(n)}$ can receive at least $E$ amount of entanglement by following the protocol in context.}

\textbf{Proof.}
To examine this, we show that by utilizing at least one particular pair of local operations it is always possible for a bipartite separable system to become entangled by extracting some fraction of entanglement from an entangled state (say the parent state), not necessarily maximally entangled, without making the parent state separable. So that, in such a scenario, the parent state can again be used to transfer entanglement to the next pair of separable states and the process can be continued. To prove this, we will consider unitaries of the same form $$U_{CA}^{(n)}=U_{BD}^{(n)}=\exp(-iHt),$$ where $H$ is expressed in Eq. \eqref{eq5}. We begin by defining the shared state of Alice and Bob after $n$th interaction as 
\begin{equation}
    \rho^{(n)}_{AB}=p^{(n)} \ket{\phi}\bra{\phi}+\left(1-p^{(n)}\right)I_D^{(n)}, \label{eq4}
\end{equation}
where $I_D^{(n)}$ is a diagonal matrix with the diagonal elements $a_1^{(n)}$, $a_2^{(n)}$, $a_3^{(n)}$, $a_4^{(n)}$. Though in general $p^{(n)}$ can be any real number for which $\rho_{AB}^{(n)}$ represent a valid state, we restrict $p^{(n)}$ to be limited within the range $0\leq p^{(n)}\leq 1$.

For $n=0$, $p^{(0)}$ can be chosen as unity. For other positive integer values of $n$, it can be easily checked that because of the particular considered form of $\rho_{AB}^{(n)}$, though after interaction with each pair of Charu$^{(n+1)}$ and Debu$^{(n+1)}$, the values of the variables $p^{(n)}$, $a_1^{(n)}$, $a_2^{(n)}$, $a_3^{(n)}$, and $a_4^{(n)}$ get changed, but the overall form of the state $\rho_{AB}^{(n)}$ remains same. 

Since $\rho^{(n)}_{AB}$ denotes a valid state, thus from the positivity, and normalization condition (due to the particular considered form of the state, the hermiticity condition is automatically satisfied) we get
\begin{eqnarray}\label{eq_state}
&&\left(a^{(n)}_1+a^{(n)}_2+a^{(n)}_3+a^{(n)}_4\right) = 1 \nonumber \\
&&a^{(n)}_2 \geq  0 \nonumber \\
&&a^{(n)}_3 \geq  0 \nonumber \\
&&2\left({p^{(n)}-1}\right)a^{(n)}_1a^{(n)}_4 \geq  p^{(n)}\left(a^{(n)}_1+a^{(n)}_4\right). \label{eq_cond_0ent}
\end{eqnarray}

We know, in the initial round, $n=0$, the parent state $\rho_{AB}^{(0)}$ is maximally entangled. At that moment, Charu$^{(1)}$ and Debu$^{(1)}$ comes and interact with Alice and Bob respectively. The interaction process is described by the unitary operations $U^{(1)}_{CA}$ and $U^{(1)}_{BD}$. The final state after the interaction can be found using Eq. \eqref{eq6}. We have observed in Sec. \ref{sec3}, that this type of interaction can transfer entanglement between Charu$^{(1)}$ and Debu$^{(1)}$ and the amount of transferred entanglement is expressed in Eq. \eqref{eq2}. Thus we can prove the theorem using mathematical induction. We assume that 
\begin{itemize}

    \item $\rho^{(n+1)}_{CD}$ is an entangled state, which one gets by applying local operation of the form expressed in Eq. \eqref{eq3} on an entangled $\rho^{(n)}_{AB}$ of the form of Eq. \eqref{eq4}.
\end{itemize}
We want to prove
\begin{enumerate}
    \item 
    \label{enu1}
After the $n$th interaction, the parent state, $\rho^{(n)}_{AB}$, transforms to a new parent state, $\rho^{(n+1)}_{AB}$, which can still be expressed in the form of Eq. \eqref{eq4} with the old variables replaced by the new ones, i.e., $p^{(n)}\rightarrow p^{(n+1)}$ and $a_i^{(n)}\rightarrow a_i^{(n+1)}$ for all $i\in \{1,2,3,4\}$. 
\item 
\label{enu2}
When a new pair, Charu$^{(n+2)}$ and Debu$^{(n+2)}$, interacts with the parent state, $\rho^{(n+1)}_{AB}$, the final state after interaction, $\rho_{CD}^{(n+2)}$ becomes entangled. 
\end{enumerate}
Since the initial parent state, $\rho^{(n)}_{AB}$, is considered to be entangled, using the positive partial transposition criteria, we get
\begin{eqnarray}\label{eq_cond_1ent}
a^{(n)}_2 a^{(n)}_3 < \left[\frac{p^{(n)}}{2\left(1-p^{(n)}\right)}\right]^2. 
\end{eqnarray}
Through the performance of the protocol $\rho_{AB}^{(n)}$ would be transformed to $\rho_{AB}^{(n+1)}=\Tr_{CD}{\left(\rho'_{CABD}\right)}$. Expressing $\rho_{AB}^{(n+1)}$ in the form presented in Eq. \eqref{eq4}, we can find the following recurrence relation on the parameters 
\begin{widetext}
\begin{eqnarray} \label{eq_a}
    &&a_1^{(n+1)}q^{(n+1)}+\frac{p^{(n+1)}}{2}=a_1^{(n)}q^{(n)}+\frac{p^{(n)}}{2}+\left(a_2^{(n)}+a_3^{(n)}\right)q^{(n)}\sin^2(2t)+\left[a_4^{(n)}q^{(n)}+\frac{p^{(n)}}{2}\right]\sin^4(2t),\label{in1}\\ 
    &&a_4^{(n+1)}q^{(n+1)}+\frac{p^{(n+1)}}{2}=\left(a_4^{(n)}q^{(n)}+\frac{p^{(n)}}{2}\right)\cos^4(2t),\label{in2}\\
    &&a_2^{(n+1)}q^{(n+1)}=\left[a_2^{(n)}q^{(n)}+\left(a_4^{(n)}q^{(n)}+\frac{p^{(n)}}{2}\right)\sin^2(2t)\right]\cos^2(2t),\label{in3}\\
    &&a_3^{(n+1)}q^{(n+1)}=\left[a_3^{(n)}q^{(n)}+\left(a_4^{(n)}q^{(n)}+\frac{p^{(n)}}{2}\right)\sin^2(2t)\right]\cos^2(2t),~~~p^{(n+1)}=p^{(n)}\cos^2(2t),\label{in4}
\end{eqnarray}
\end{widetext}
where, $q^{(n)}=1-p^{(n)}$. Let us use the notation, $X^{(n)}=a_4^{(n)}q^{(n)}+\frac{p^{(n)}}{2}$. From Eq. \eqref{in2}, we see if $X^{(n)}<0$ then $X^{(n+1)}<0$ for all $n$ and all $\cos(2t)\neq 0$. Similarly, $X^{(n)}>0$ implies 
$X^{(n+1)}>0$ for all $n$ and $\cos(2t)\neq 0$. Since we have considered $\rho^{(0)}=\ket{\phi^+}\bra{\phi^+}$, that is, $p^{(0)}=1$, thus we have $X^{(0)}>0$. Hence, $X^{(n)}>0$, for all $n$. Thus we have
\begin{equation}
    2a_4^{(n)}+\frac{p^{(n)}}{1-p^{(n)}}>0~\text{for all }n. \label{new1}
\end{equation}
We realize $\rho_{AB}^{(n)}$ can be mapped to $\rho_{AB}^{(n+1)}$ just by mapping the individual parameters, $a^{(n)}_i\rightarrow a_i^{(n+1)}$ and $p^{(n)}\rightarrow p^{(n+1)}$. Therefore, \ref{enu1} has been proved.

Next we want to prove that $\rho_{CD}^{(n+2)}$ is entangled. Since, we have considered $\rho_{CD}^{(n+1)}$ to be entangled, therefore, using the PPT criteria, which provides a necessary and sufficient condition for sharing entanglement, we have 
\begin{eqnarray}
    \left[2a^{(n)}_2+\chi^{(n)} \right] \left[2a^{(n)}_3+\chi^{(n)} \right] 
    < \left(\frac{p^{(n)}}{(1-p^{(n)})}\right)^2\label{eq_path1_chi1}
    \nonumber\\ \text{and}~\sin(2t)\neq 0
\end{eqnarray}
where $\chi^{(n)}=\left[2a_4^{(n)}+\frac{p^{(n)}}{1-p^{(n)}}\right]\cos^2(2t) $. From inequaulity \eqref{new1}, it is clear that $\chi^{(n)}$ is positive for all $n$ and $\cos(2t)\neq 0$. Now in order to have an entangled $\rho_{CD}^{(n+2)}$, the following condition must be satisfied which is obtained using the PPT criterion and the recursion relations for the state parameters, 
\begin{eqnarray}\label{eq_path1_chi2}
    \left[2a^{(n)}_2+\xi^{(n)} \right] \left[2a^{(n)}_3+\xi^{(n)} \right]  < \left(\frac{p^{(n)}}{(1-p^{(n)})}\right)^2, \label{eq7}
\end{eqnarray}
where 
\begin{eqnarray}
    \xi^{(n)}=\chi^{(n)}\frac{(1-\cos^2(2t)\sin^2(2t))}{\cos^2(2t)}.\nonumber
\end{eqnarray}
In can be seen that by replacing $\xi^{(n)}$ in place of $\chi^{(n)}$ in inequality \eqref{eq_path1_chi1} provides the condition expressed in \eqref{eq7}. One can easily see that $\xi^{(n)}\geq \chi^{(n)}$ and the gap between $\xi^{(n)}$ and $\chi^{(n)}$ can be made arbitrarily small for the smaller values of the parameter $t$. And for $t=0$, $\xi^{(n)}= \chi^{(n)}$.  Now, since there is a gap between $f(\chi^{(n)})=\left[2a^{(n)}_2+\chi^{(n)} \right] \left[2a^{(n)}_3+\chi^{(n)} \right]$ and $\left(\frac{p^{(n)}}{(1-p^{(n)})}\right)^2$, i.e., $f(\chi^{(n)})-\left(\frac{p^{(n)}}{(1-p^{(n)})}\right)^2<0$, though $\xi^{(n)}>\chi^{(n)}$, we can always choose a non-zero but arbitrarily small enough $t$ such that the inequality $f(\xi^{(n)})-\left(\frac{p^{(n)}}{(1-p^{(n)})}\right)^2<0$ holds. This proves that we can choose $t$ in such a way that $\rho_{CD}^{(n+2)}$ be entangled, given $\rho_{CD}^{(n+1)}$ is entangled.
 
 
The satisfaction of the condition for $\rho_{CD}^{(n+2)}$ to be entangled automatically implies $\rho_{AB}^{n+1}$ is also entangled otherwise where do have Charu$^{(n+2)}$ and Debu$^{(n+2)}$ received that entanglement! Thus it completes the proof with $E$ given by $\min_n E_{CD}^{(n)}$. It can be observed that the range of $t$ for which the entanglement transfer protocol remains valid decreases with the increase of $N$, making $\lambda$ smaller. So it becomes more and more difficult to transfer entanglement to a larger number of pairs, additionally, each pair involved in such processes receives a reduced amount of entanglement. 
\hfill \(\blacksquare\)
\vspace{3mm}\\
\textbf{Remark:} In the previous proof we took $p^{(0)}=1$ and thus we got $X^{(n)}>0$. But even if instead of a maximally entangled state another $\rho_{AB}^{(0)}$ is considered such that $X^{(0)}$ become negative, i.e., $a_4^{(0)}<\frac{p^{(0)}}{2(p^{(0)}-1)}$, the sequential entanglement transfer may still be possible.
In this case, with the consideration that $\rho_{AB}^{(n)}$ is entangled, the criterion for $\rho_{CD}^{(n+1)}$ to be entangled reduces to 
\begin{eqnarray} \label{eq_cond_3ent}
    a_4^{(n)} &<& \frac{p^{(n)}}{2(p^{(n)}-1)},\label{eq8}
\end{eqnarray}
where $p^{(n+1)}=p^{(n)}\cos^2(2t)$. To continue the process, $\rho_{AB}^{(n+1)}$ should also be entangled so that it can transfer some of the entanglement to the next pair of Charu and Debu. The conditions for $\rho_{AB}^{(n+1)}$ to be entangled are given by
\begin{eqnarray} \label{eq_cond_3ent}
    a_4^{(n)} &<& \frac{p^{(n)}}{2\left(p^{(n)}-1\right)},\label{eq9}\\
     \sin^2(2t) &<& \frac{1-p^{(n)}}{|a^{(n)}_4(1-p^{(n)})+p^{(n)}/2|}\nonumber\\&&\left(a^{(n)}_2+a^{(n)}_3\right).\label{eq99c}
\end{eqnarray}
Conditions expressed in \eqref{eq8} and \eqref{eq9} are the same. Since $X^{(n)}<0$ confirms $X^{(n+1)}<0$, and initially we considered $X^{(0)}<0$, it automatically proves that inequality \eqref{eq8} or \eqref{eq9} would be satisfied for all $n$. Moreover, we can always choose $t$ in such a way that inequality \eqref{eq99c} also be satisfied for the desired number of rounds, $n$.

\section{Conclusion}
\label{harkin-choT}
Entanglement, a fundamental and invaluable concept in quantum mechanics, offers numerous key advantages and applications. This paper focuses on the scenario where a pair of initially separable states can become entangled through 
local
interactions with each of the parties 
of
an entangled pair. In this context, Charu and Debu independently engage with Alice and Bob, respectively, who share a maximally entangled state. The 
investigation centers around the feasibility of transferring entanglement from the Alice-Bob pair to the Charu-Debu pair using local joint unitary operations involving Alice-Charu and Bob-Debu. Expanding upon this concept, we explore the sequential transfer of entanglement to multiple pairs of Charus and Debus. By systematically optimizing all potential local joint unitaries, we examine the maximum number of pairs to which a fixed amount of entanglement can be transferred.  Interestingly, our findings reveal the existence of at least one pair of unitaries capable of transferring a nonzero amount of entanglement to an arbitrarily large number of pairs. We also analyzed the amount of entanglement that can be transferred using such unitaries, to a fixed number of pairs, assuming that each pair extract the same amount of entanglement. We also examined to how many pairs, a fixed amount of entanglement can be transferred sequentially using such unitaries.
 \acknowledgements 
The research of KS was supported in part by the INFOSYS scholarship. C. S. acknowledges funding from the QuantERA II Programme that has received funding from the European Union’s Horizon 2020 research and innovation programme under Grant Agreement No 101017733 and the F.R.S-FNRS Pint-Multi programme under Grant Agreement R.8014.21,
from the European Union’s Horizon Europe research and innovation programme under the project "Quantum Security Networks Partnership" (QSNP, grant agreement No 101114043), from the F.R.S-FNRS through the PDR T.0171.22, from the FWO and F.R.S.-FNRS under the Excellence of Science (EOS) programme project 40007526, from the FWO through the BeQuNet SBO project S008323N. Funded by the European Union. Views and opinions expressed are however those of the authors only and do not necessarily reflect those of the European Union, who cannot be held responsible for them. We acknowledge partial support from the Department of Science and Technology, Government of India, through the QuEST grant (grant number DST/ICPS/QUST/Theme-3/2019/120). 
\bibliography{ref}

\begin{thebibliography}{26}%
\makeatletter
\providecommand \@ifxundefined [1]{%
 \@ifx{#1\undefined}
}%
\providecommand \@ifnum [1]{%
 \ifnum #1\expandafter \@firstoftwo
 \else \expandafter \@secondoftwo
 \fi
}%
\providecommand \@ifx [1]{%
 \ifx #1\expandafter \@firstoftwo
 \else \expandafter \@secondoftwo
 \fi
}%
\providecommand \natexlab [1]{#1}%
\providecommand \enquote  [1]{``#1''}%
\providecommand \bibnamefont  [1]{#1}%
\providecommand \bibfnamefont [1]{#1}%
\providecommand \citenamefont [1]{#1}%
\providecommand \href@noop [0]{\@secondoftwo}%
\providecommand \href [0]{\begingroup \@sanitize@url \@href}%
\providecommand \@href[1]{\@@startlink{#1}\@@href}%
\providecommand \@@href[1]{\endgroup#1\@@endlink}%
\providecommand \@sanitize@url [0]{\catcode `\\12\catcode `\$12\catcode
  `\&12\catcode `\#12\catcode `\^12\catcode `\_12\catcode `\%12\relax}%
\providecommand \@@startlink[1]{}%
\providecommand \@@endlink[0]{}%
\providecommand \url  [0]{\begingroup\@sanitize@url \@url }%
\providecommand \@url [1]{\endgroup\@href {#1}{\urlprefix }}%
\providecommand \urlprefix  [0]{URL }%
\providecommand \Eprint [0]{\href }%
\providecommand \doibase [0]{http://dx.doi.org/}%
\providecommand \selectlanguage [0]{\@gobble}%
\providecommand \bibinfo  [0]{\@secondoftwo}%
\providecommand \bibfield  [0]{\@secondoftwo}%
\providecommand \translation [1]{[#1]}%
\providecommand \BibitemOpen [0]{}%
\providecommand \bibitemStop [0]{}%
\providecommand \bibitemNoStop [0]{.\EOS\space}%
\providecommand \EOS [0]{\spacefactor3000\relax}%
\providecommand \BibitemShut  [1]{\csname bibitem#1\endcsname}%
\let\auto@bib@innerbib\@empty
\bibitem [{\citenamefont {Einstein}\ \emph {et~al.}(1935)\citenamefont
  {Einstein}, \citenamefont {Podolsky},\ and\ \citenamefont {Rosen}}]{EPR1}%
  \BibitemOpen
  \bibfield  {author} {\bibinfo {author} {\bibfnamefont {A.}~\bibnamefont
  {Einstein}}, \bibinfo {author} {\bibfnamefont {B.}~\bibnamefont {Podolsky}},
  \ and\ \bibinfo {author} {\bibfnamefont {N.}~\bibnamefont {Rosen}},\
  }\bibfield  {title} {\enquote {\bibinfo {title} {Can quantum-mechanical
  description of physical reality be considered complete?}}\ }\href
  {https://link.aps.org/doi/10.1103/PhysRev.47.777} {\bibfield  {journal}
  {\bibinfo  {journal} {Phys. Rev.}\ }\textbf {\bibinfo {volume} {47}},\
  \bibinfo {pages} {777} (\bibinfo {year} {1935})}\BibitemShut {NoStop}%
\bibitem [{\citenamefont {Horodecki}\ \emph {et~al.}(2009)\citenamefont
  {Horodecki}, \citenamefont {Horodecki}, \citenamefont {Horodecki},\ and\
  \citenamefont {Horodecki}}]{ent1}%
  \BibitemOpen
  \bibfield  {author} {\bibinfo {author} {\bibfnamefont {R.}~\bibnamefont
  {Horodecki}}, \bibinfo {author} {\bibfnamefont {P.}~\bibnamefont
  {Horodecki}}, \bibinfo {author} {\bibfnamefont {M.}~\bibnamefont
  {Horodecki}}, \ and\ \bibinfo {author} {\bibfnamefont {K.}~\bibnamefont
  {Horodecki}},\ }\bibfield  {title} {\enquote {\bibinfo {title} {Quantum
  entanglement},}\ }\href {https://link.aps.org/doi/10.1103/RevModPhys.81.865}
  {\bibfield  {journal} {\bibinfo  {journal} {Rev. Mod. Phys.}\ }\textbf
  {\bibinfo {volume} {81}},\ \bibinfo {pages} {865} (\bibinfo {year}
  {2009})}\BibitemShut {NoStop}%
\bibitem [{\citenamefont {Das}\ \emph {et~al.}(2017)\citenamefont {Das},
  \citenamefont {Chanda}, \citenamefont {Lewenstein}, \citenamefont {Sanpera},
  \citenamefont {De},\ and\ \citenamefont {Sen}}]{ent2}%
  \BibitemOpen
  \bibfield  {author} {\bibinfo {author} {\bibfnamefont {S.}~\bibnamefont
  {Das}}, \bibinfo {author} {\bibfnamefont {T.}~\bibnamefont {Chanda}},
  \bibinfo {author} {\bibfnamefont {M.}~\bibnamefont {Lewenstein}}, \bibinfo
  {author} {\bibfnamefont {A.}~\bibnamefont {Sanpera}}, \bibinfo {author}
  {\bibfnamefont {A.~S.}\ \bibnamefont {De}}, \ and\ \bibinfo {author}
  {\bibfnamefont {U.}~\bibnamefont {Sen}},\ }\bibfield  {title} {\enquote
  {\bibinfo {title} {The separability versus entanglement problem},}\ }\href
  {https://arxiv.org/abs/1701.02187} {\bibfield  {journal} {\bibinfo  {journal}
  {arXiv:1701.02187}\ } (\bibinfo {year} {2017})}\BibitemShut {NoStop}%
\bibitem [{\citenamefont {Bennett}\ \emph {et~al.}(1993)\citenamefont
  {Bennett}, \citenamefont {Brassard}, \citenamefont {Cr\'epeau}, \citenamefont
  {Jozsa}, \citenamefont {Peres},\ and\ \citenamefont
  {Wootters}}]{Teleportation}%
  \BibitemOpen
  \bibfield  {author} {\bibinfo {author} {\bibfnamefont {C.~H.}\ \bibnamefont
  {Bennett}}, \bibinfo {author} {\bibfnamefont {G.}~\bibnamefont {Brassard}},
  \bibinfo {author} {\bibfnamefont {C.}~\bibnamefont {Cr\'epeau}}, \bibinfo
  {author} {\bibfnamefont {R.}~\bibnamefont {Jozsa}}, \bibinfo {author}
  {\bibfnamefont {A.}~\bibnamefont {Peres}}, \ and\ \bibinfo {author}
  {\bibfnamefont {W.~K.}\ \bibnamefont {Wootters}},\ }\bibfield  {title}
  {\enquote {\bibinfo {title} {Teleporting an unknown quantum state via dual
  classical and einstein-podolsky-rosen channels},}\ }\href
  {https://link.aps.org/doi/10.1103/PhysRevLett.70.1895} {\bibfield  {journal}
  {\bibinfo  {journal} {Phys. Rev. Lett.}\ }\textbf {\bibinfo {volume} {70}},\
  \bibinfo {pages} {1895} (\bibinfo {year} {1993})}\BibitemShut {NoStop}%
\bibitem [{\citenamefont {Bennett}\ and\ \citenamefont
  {Wiesner}(1992)}]{Densed}%
  \BibitemOpen
  \bibfield  {author} {\bibinfo {author} {\bibfnamefont {C.~H.}\ \bibnamefont
  {Bennett}}\ and\ \bibinfo {author} {\bibfnamefont {S.~J.}\ \bibnamefont
  {Wiesner}},\ }\bibfield  {title} {\enquote {\bibinfo {title} {Communication
  via one- and two-particle operators on einstein-podolsky-rosen states},}\
  }\href {https://link.aps.org/doi/10.1103/PhysRevLett.69.2881} {\bibfield
  {journal} {\bibinfo  {journal} {Phys. Rev. Lett.}\ }\textbf {\bibinfo
  {volume} {69}},\ \bibinfo {pages} {2881} (\bibinfo {year}
  {1992})}\BibitemShut {NoStop}%
\bibitem [{\citenamefont {Gisin}\ \emph {et~al.}(2002)\citenamefont {Gisin},
  \citenamefont {Ribordy}, \citenamefont {Tittel},\ and\ \citenamefont
  {Zbinden}}]{crypto1}%
  \BibitemOpen
  \bibfield  {author} {\bibinfo {author} {\bibfnamefont {N.}~\bibnamefont
  {Gisin}}, \bibinfo {author} {\bibfnamefont {G.}~\bibnamefont {Ribordy}},
  \bibinfo {author} {\bibfnamefont {W.}~\bibnamefont {Tittel}}, \ and\ \bibinfo
  {author} {\bibfnamefont {H.}~\bibnamefont {Zbinden}},\ }\bibfield  {title}
  {\enquote {\bibinfo {title} {Quantum cryptography},}\ }\href {\doibase
  10.1103/revmodphys.74.145} {\bibfield  {journal} {\bibinfo  {journal}
  {Reviews of Modern Physics}\ }\textbf {\bibinfo {volume} {74}},\ \bibinfo
  {pages} {145--195} (\bibinfo {year} {2002})}\BibitemShut {NoStop}%
\bibitem [{\citenamefont {Ekert}(1991)}]{crypto2}%
  \BibitemOpen
  \bibfield  {author} {\bibinfo {author} {\bibfnamefont {A.~K.}\ \bibnamefont
  {Ekert}},\ }\bibfield  {title} {\enquote {\bibinfo {title} {Quantum
  cryptography},}\ }\href
  {https://journals.aps.org/prl/abstract/10.1103/PhysRevLett.67.661%2Frevmodphys.74.145}
  {\bibfield  {journal} {\bibinfo  {journal} {Phys. Rev. Lett.}\ }\textbf
  {\bibinfo {volume} {67}},\ \bibinfo {pages} {661} (\bibinfo {year}
  {1991})}\BibitemShut {NoStop}%
\bibitem [{\citenamefont {Bose}\ \emph {et~al.}(1998)\citenamefont {Bose},
  \citenamefont {Vedral},\ and\ \citenamefont {Knight}}]{swap1}%
  \BibitemOpen
  \bibfield  {author} {\bibinfo {author} {\bibfnamefont {S.}~\bibnamefont
  {Bose}}, \bibinfo {author} {\bibfnamefont {V.}~\bibnamefont {Vedral}}, \ and\
  \bibinfo {author} {\bibfnamefont {P.~L.}\ \bibnamefont {Knight}},\ }\bibfield
   {title} {\enquote {\bibinfo {title} {Multiparticle generalization of
  entanglement swapping},}\ }\href {\doibase 10.1103/physreva.57.822}
  {\bibfield  {journal} {\bibinfo  {journal} {Phys. Rev. A}\ }\textbf {\bibinfo
  {volume} {57}},\ \bibinfo {pages} {822} (\bibinfo {year} {1998})}\BibitemShut
  {NoStop}%
\bibitem [{\citenamefont {\ifmmode~\dot{Z}\else \.{Z}\fi{}ukowski}\ \emph
  {et~al.}(1993)\citenamefont {\ifmmode~\dot{Z}\else \.{Z}\fi{}ukowski},
  \citenamefont {Zeilinger}, \citenamefont {Horne},\ and\ \citenamefont
  {Ekert}}]{swap2}%
  \BibitemOpen
  \bibfield  {author} {\bibinfo {author} {\bibfnamefont {M.}~\bibnamefont
  {\ifmmode~\dot{Z}\else \.{Z}\fi{}ukowski}}, \bibinfo {author} {\bibfnamefont
  {A.}~\bibnamefont {Zeilinger}}, \bibinfo {author} {\bibfnamefont {M.~A.}\
  \bibnamefont {Horne}}, \ and\ \bibinfo {author} {\bibfnamefont {A.~K.}\
  \bibnamefont {Ekert}},\ }\bibfield  {title} {\enquote {\bibinfo {title}
  {``event-ready-detectors'' bell experiment via entanglement swapping},}\
  }\href {\doibase 10.1103/PhysRevLett.71.4287} {\bibfield  {journal} {\bibinfo
   {journal} {Phys. Rev. Lett.}\ }\textbf {\bibinfo {volume} {71}},\ \bibinfo
  {pages} {4287} (\bibinfo {year} {1993})}\BibitemShut {NoStop}%
\bibitem [{\citenamefont {Colbeck}\ and\ \citenamefont
  {Renner}(2012)}]{randomness}%
  \BibitemOpen
  \bibfield  {author} {\bibinfo {author} {\bibfnamefont {R.}~\bibnamefont
  {Colbeck}}\ and\ \bibinfo {author} {\bibfnamefont {R.}~\bibnamefont
  {Renner}},\ }\bibfield  {title} {\enquote {\bibinfo {title} {Free randomness
  can be amplified},}\ }\href {\doibase 10.1038/nphys2300} {\bibfield
  {journal} {\bibinfo  {journal} {Nature Physics}\ }\textbf {\bibinfo {volume}
  {8}},\ \bibinfo {pages} {450} (\bibinfo {year} {2012})}\BibitemShut {NoStop}%
\bibitem [{\citenamefont {Silva}\ \emph {et~al.}(2015)\citenamefont {Silva},
  \citenamefont {Gisin}, \citenamefont {Guryanova},\ and\ \citenamefont
  {Popescu}}]{Silva15}%
  \BibitemOpen
  \bibfield  {author} {\bibinfo {author} {\bibfnamefont {R.}~\bibnamefont
  {Silva}}, \bibinfo {author} {\bibfnamefont {N.}~\bibnamefont {Gisin}},
  \bibinfo {author} {\bibfnamefont {Y.}~\bibnamefont {Guryanova}}, \ and\
  \bibinfo {author} {\bibfnamefont {S.}~\bibnamefont {Popescu}},\ }\bibfield
  {title} {\enquote {\bibinfo {title} {Multiple observers can share the
  nonlocality of half of an entangled pair by using optimal weak
  measurements},}\ }\href {\doibase 10.1103/PhysRevLett.114.250401} {\bibfield
  {journal} {\bibinfo  {journal} {Phys. Rev. Lett.}\ }\textbf {\bibinfo
  {volume} {114}},\ \bibinfo {pages} {250401} (\bibinfo {year}
  {2015})}\BibitemShut {NoStop}%
\bibitem [{\citenamefont {Mal}\ \emph {et~al.}(2016)\citenamefont {Mal},
  \citenamefont {Majumdar},\ and\ \citenamefont {Home}}]{Mal16}%
  \BibitemOpen
  \bibfield  {author} {\bibinfo {author} {\bibfnamefont {S.}~\bibnamefont
  {Mal}}, \bibinfo {author} {\bibfnamefont {A.~S.}\ \bibnamefont {Majumdar}}, \
  and\ \bibinfo {author} {\bibfnamefont {D.}~\bibnamefont {Home}},\ }\bibfield
  {title} {\enquote {\bibinfo {title} {Sharing of nonlocality of a single
  member of an entangled pair of qubits is not possible by more than two
  unbiased observers on the other wing},}\ }\href {\doibase
  10.3390/math4030048} {\bibfield  {journal} {\bibinfo  {journal}
  {Mathematics}\ }\textbf {\bibinfo {volume} {4}} (\bibinfo {year} {2016}),\
  10.3390/math4030048}\BibitemShut {NoStop}%
\bibitem [{\citenamefont {Saha}\ \emph {et~al.}(2019)\citenamefont {Saha},
  \citenamefont {Das}, \citenamefont {Sasmal}, \citenamefont {Sarkar},
  \citenamefont {Mukherjee}, \citenamefont {Roy},\ and\ \citenamefont
  {Bhattacharya}}]{Saha19}%
  \BibitemOpen
  \bibfield  {author} {\bibinfo {author} {\bibfnamefont {S.}~\bibnamefont
  {Saha}}, \bibinfo {author} {\bibfnamefont {D.}~\bibnamefont {Das}}, \bibinfo
  {author} {\bibfnamefont {S.}~\bibnamefont {Sasmal}}, \bibinfo {author}
  {\bibfnamefont {D.}~\bibnamefont {Sarkar}}, \bibinfo {author} {\bibfnamefont
  {K.}~\bibnamefont {Mukherjee}}, \bibinfo {author} {\bibfnamefont
  {A.}~\bibnamefont {Roy}}, \ and\ \bibinfo {author} {\bibfnamefont {S.~S.}\
  \bibnamefont {Bhattacharya}},\ }\bibfield  {title} {\enquote {\bibinfo
  {title} {Sharing of tripartite nonlocality by multiple observers measuring
  sequentially at one side},}\ }\href {\doibase 10.1007/s11128-018-2161-x}
  {\bibfield  {journal} {\bibinfo  {journal} {Quantum Information Processing}\
  }\textbf {\bibinfo {volume} {18}},\ \bibinfo {pages} {42} (\bibinfo {year}
  {2019})}\BibitemShut {NoStop}%
\bibitem [{\citenamefont {Das}\ \emph {et~al.}(2019)\citenamefont {Das},
  \citenamefont {Ghosal}, \citenamefont {Sasmal}, \citenamefont {Mal},\ and\
  \citenamefont {Majumdar}}]{Das19}%
  \BibitemOpen
  \bibfield  {author} {\bibinfo {author} {\bibfnamefont {D.}~\bibnamefont
  {Das}}, \bibinfo {author} {\bibfnamefont {A.}~\bibnamefont {Ghosal}},
  \bibinfo {author} {\bibfnamefont {S.}~\bibnamefont {Sasmal}}, \bibinfo
  {author} {\bibfnamefont {S.}~\bibnamefont {Mal}}, \ and\ \bibinfo {author}
  {\bibfnamefont {A.~S.}\ \bibnamefont {Majumdar}},\ }\bibfield  {title}
  {\enquote {\bibinfo {title} {Facets of bipartite nonlocality sharing by
  multiple observers via sequential measurements},}\ }\href {\doibase
  10.1103/PhysRevA.99.022305} {\bibfield  {journal} {\bibinfo  {journal} {Phys.
  Rev. A}\ }\textbf {\bibinfo {volume} {99}},\ \bibinfo {pages} {022305}
  (\bibinfo {year} {2019})}\BibitemShut {NoStop}%
\bibitem [{\citenamefont {Brown}\ and\ \citenamefont
  {Colbeck}(2020)}]{Brown20}%
  \BibitemOpen
  \bibfield  {author} {\bibinfo {author} {\bibfnamefont {P.~J.}\ \bibnamefont
  {Brown}}\ and\ \bibinfo {author} {\bibfnamefont {R.}~\bibnamefont
  {Colbeck}},\ }\bibfield  {title} {\enquote {\bibinfo {title} {Arbitrarily
  many independent observers can share the nonlocality of a single maximally
  entangled qubit pair},}\ }\href {\doibase 10.1103/PhysRevLett.125.090401}
  {\bibfield  {journal} {\bibinfo  {journal} {Phys. Rev. Lett.}\ }\textbf
  {\bibinfo {volume} {125}},\ \bibinfo {pages} {090401} (\bibinfo {year}
  {2020})}\BibitemShut {NoStop}%
\bibitem [{\citenamefont {Cabello}(2021)}]{cabello20}%
  \BibitemOpen
  \bibfield  {author} {\bibinfo {author} {\bibfnamefont {A.}~\bibnamefont
  {Cabello}},\ }\href@noop {} {\enquote {\bibinfo {title} {Bell nonlocality
  between sequential pairs of observers},}\ } (\bibinfo {year} {2021}),\
  \Eprint {http://arxiv.org/abs/2103.11844} {arXiv:2103.11844 [quant-ph]}
  \BibitemShut {NoStop}%
\bibitem [{\citenamefont {Zhang}\ and\ \citenamefont {Fei}(2021)}]{Fei21}%
  \BibitemOpen
  \bibfield  {author} {\bibinfo {author} {\bibfnamefont {T.}~\bibnamefont
  {Zhang}}\ and\ \bibinfo {author} {\bibfnamefont {S.-M.}\ \bibnamefont
  {Fei}},\ }\bibfield  {title} {\enquote {\bibinfo {title} {Sharing quantum
  nonlocality and genuine nonlocality with independent observables},}\ }\href
  {\doibase 10.1103/PhysRevA.103.032216} {\bibfield  {journal} {\bibinfo
  {journal} {Phys. Rev. A}\ }\textbf {\bibinfo {volume} {103}},\ \bibinfo
  {pages} {032216} (\bibinfo {year} {2021})}\BibitemShut {NoStop}%
\bibitem [{\citenamefont {Bera}\ \emph {et~al.}(2018)\citenamefont {Bera},
  \citenamefont {Mal}, \citenamefont {Sen(De)},\ and\ \citenamefont
  {Sen}}]{Bera18}%
  \BibitemOpen
  \bibfield  {author} {\bibinfo {author} {\bibfnamefont {A.}~\bibnamefont
  {Bera}}, \bibinfo {author} {\bibfnamefont {S.}~\bibnamefont {Mal}}, \bibinfo
  {author} {\bibfnamefont {A.}~\bibnamefont {Sen(De)}}, \ and\ \bibinfo
  {author} {\bibfnamefont {U.}~\bibnamefont {Sen}},\ }\bibfield  {title}
  {\enquote {\bibinfo {title} {Witnessing bipartite entanglement sequentially
  by multiple observers},}\ }\href {\doibase 10.1103/PhysRevA.98.062304}
  {\bibfield  {journal} {\bibinfo  {journal} {Phys. Rev. A}\ }\textbf {\bibinfo
  {volume} {98}},\ \bibinfo {pages} {062304} (\bibinfo {year}
  {2018})}\BibitemShut {NoStop}%
\bibitem [{\citenamefont {Srivastava}\ \emph {et~al.}(2021)\citenamefont
  {Srivastava}, \citenamefont {Mal}, \citenamefont {Sen(De)},\ and\
  \citenamefont {Sen}}]{sriv21}%
  \BibitemOpen
  \bibfield  {author} {\bibinfo {author} {\bibfnamefont {C.}~\bibnamefont
  {Srivastava}}, \bibinfo {author} {\bibfnamefont {S.}~\bibnamefont {Mal}},
  \bibinfo {author} {\bibfnamefont {A.}~\bibnamefont {Sen(De)}}, \ and\
  \bibinfo {author} {\bibfnamefont {U.}~\bibnamefont {Sen}},\ }\bibfield
  {title} {\enquote {\bibinfo {title} {Sequential
  measurement-device-independent entanglement detection by multiple
  observers},}\ }\href {\doibase 10.1103/PhysRevA.103.032408} {\bibfield
  {journal} {\bibinfo  {journal} {Phys. Rev. A}\ }\textbf {\bibinfo {volume}
  {103}},\ \bibinfo {pages} {032408} (\bibinfo {year} {2021})}\BibitemShut
  {NoStop}%
\bibitem [{\citenamefont {Srivastava}\ \emph
  {et~al.}(2022{\natexlab{a}})\citenamefont {Srivastava}, \citenamefont
  {Pandit},\ and\ \citenamefont {Sen}}]{Srivastava22}%
  \BibitemOpen
  \bibfield  {author} {\bibinfo {author} {\bibfnamefont {C.}~\bibnamefont
  {Srivastava}}, \bibinfo {author} {\bibfnamefont {M.}~\bibnamefont {Pandit}},
  \ and\ \bibinfo {author} {\bibfnamefont {U.}~\bibnamefont {Sen}},\ }\bibfield
   {title} {\enquote {\bibinfo {title} {Entanglement witnessing by arbitrarily
  many independent observers recycling a local quantum shared state},}\ }\href
  {\doibase 10.1103/PhysRevA.105.062413} {\bibfield  {journal} {\bibinfo
  {journal} {Phys. Rev. A}\ }\textbf {\bibinfo {volume} {105}},\ \bibinfo
  {pages} {062413} (\bibinfo {year} {2022}{\natexlab{a}})}\BibitemShut
  {NoStop}%
\bibitem [{\citenamefont {Pandit}\ \emph {et~al.}(2022)\citenamefont {Pandit},
  \citenamefont {Srivastava},\ and\ \citenamefont {Sen}}]{Pandit22}%
  \BibitemOpen
  \bibfield  {author} {\bibinfo {author} {\bibfnamefont {M.}~\bibnamefont
  {Pandit}}, \bibinfo {author} {\bibfnamefont {C.}~\bibnamefont {Srivastava}},
  \ and\ \bibinfo {author} {\bibfnamefont {U.}~\bibnamefont {Sen}},\ }\bibfield
   {title} {\enquote {\bibinfo {title} {Recycled entanglement detection by
  arbitrarily many sequential and independent pairs of observers},}\ }\href
  {\doibase 10.1103/PhysRevA.106.032419} {\bibfield  {journal} {\bibinfo
  {journal} {Phys. Rev. A}\ }\textbf {\bibinfo {volume} {106}},\ \bibinfo
  {pages} {032419} (\bibinfo {year} {2022})}\BibitemShut {NoStop}%
\bibitem [{\citenamefont {Srivastava}\ \emph
  {et~al.}(2022{\natexlab{b}})\citenamefont {Srivastava}, \citenamefont
  {Pandit},\ and\ \citenamefont {Sen}}]{srivastava_GME1_22}%
  \BibitemOpen
  \bibfield  {author} {\bibinfo {author} {\bibfnamefont {C.}~\bibnamefont
  {Srivastava}}, \bibinfo {author} {\bibfnamefont {M.}~\bibnamefont {Pandit}},
  \ and\ \bibinfo {author} {\bibfnamefont {U.}~\bibnamefont {Sen}},\
  }\href@noop {} {\enquote {\bibinfo {title} {Recycled detection of genuine
  multiparty entanglement of unlimitedly stretched array of parties and
  arbitrarily long series of sequential observers},}\ } (\bibinfo {year}
  {2022}{\natexlab{b}}),\ \Eprint {http://arxiv.org/abs/2205.02695}
  {arXiv:2205.02695 [quant-ph]} \BibitemShut {NoStop}%
\bibitem [{\citenamefont {Srivastava}\ \emph
  {et~al.}(2022{\natexlab{c}})\citenamefont {Srivastava}, \citenamefont
  {Pandit},\ and\ \citenamefont {Sen}}]{srivastava_GME2_22}%
  \BibitemOpen
  \bibfield  {author} {\bibinfo {author} {\bibfnamefont {C.}~\bibnamefont
  {Srivastava}}, \bibinfo {author} {\bibfnamefont {M.}~\bibnamefont {Pandit}},
  \ and\ \bibinfo {author} {\bibfnamefont {U.}~\bibnamefont {Sen}},\
  }\href@noop {} {\enquote {\bibinfo {title} {Sequential detection of genuine
  multipartite entanglement is unbounded for entire hierarchy of number of
  qubits recycled},}\ } (\bibinfo {year} {2022}{\natexlab{c}}),\ \Eprint
  {http://arxiv.org/abs/2208.08435} {arXiv:2208.08435 [quant-ph]} \BibitemShut
  {NoStop}%
\bibitem [{\citenamefont {Mao}\ \emph {et~al.}(2022)\citenamefont {Mao},
  \citenamefont {Li}, \citenamefont {Steffinlongo}, \citenamefont {Guo},
  \citenamefont {Liu}, \citenamefont {Xu}, \citenamefont {Gisin}, \citenamefont
  {Tavakoli},\ and\ \citenamefont {Fan}}]{Mao22}%
  \BibitemOpen
  \bibfield  {author} {\bibinfo {author} {\bibfnamefont {Y.-L.}\ \bibnamefont
  {Mao}}, \bibinfo {author} {\bibfnamefont {Z.-D.}\ \bibnamefont {Li}},
  \bibinfo {author} {\bibfnamefont {A.}~\bibnamefont {Steffinlongo}}, \bibinfo
  {author} {\bibfnamefont {B.}~\bibnamefont {Guo}}, \bibinfo {author}
  {\bibfnamefont {B.}~\bibnamefont {Liu}}, \bibinfo {author} {\bibfnamefont
  {S.}~\bibnamefont {Xu}}, \bibinfo {author} {\bibfnamefont {N.}~\bibnamefont
  {Gisin}}, \bibinfo {author} {\bibfnamefont {A.}~\bibnamefont {Tavakoli}}, \
  and\ \bibinfo {author} {\bibfnamefont {J.}~\bibnamefont {Fan}},\ }\href@noop
  {} {\enquote {\bibinfo {title} {Recycling nonlocality in a quantum
  network},}\ } (\bibinfo {year} {2022}),\ \Eprint
  {http://arxiv.org/abs/2202.04840} {arXiv:2202.04840 [quant-ph]} \BibitemShut
  {NoStop}%
\bibitem [{\citenamefont {Curchod}\ \emph {et~al.}(2017)\citenamefont
  {Curchod}, \citenamefont {Johansson}, \citenamefont {Augusiak}, \citenamefont
  {Hoban}, \citenamefont {Wittek},\ and\ \citenamefont
  {Ac\'{\i}n}}]{Curchod17}%
  \BibitemOpen
  \bibfield  {author} {\bibinfo {author} {\bibfnamefont {F.~J.}\ \bibnamefont
  {Curchod}}, \bibinfo {author} {\bibfnamefont {M.}~\bibnamefont {Johansson}},
  \bibinfo {author} {\bibfnamefont {R.}~\bibnamefont {Augusiak}}, \bibinfo
  {author} {\bibfnamefont {M.~J.}\ \bibnamefont {Hoban}}, \bibinfo {author}
  {\bibfnamefont {P.}~\bibnamefont {Wittek}}, \ and\ \bibinfo {author}
  {\bibfnamefont {A.}~\bibnamefont {Ac\'{\i}n}},\ }\bibfield  {title} {\enquote
  {\bibinfo {title} {Unbounded randomness certification using sequences of
  measurements},}\ }\href {\doibase 10.1103/PhysRevA.95.020102} {\bibfield
  {journal} {\bibinfo  {journal} {Phys. Rev. A}\ }\textbf {\bibinfo {volume}
  {95}},\ \bibinfo {pages} {020102} (\bibinfo {year} {2017})}\BibitemShut
  {NoStop}%
\bibitem [{\citenamefont {Bru{\ss}}(1999)}]{ent-splitting}%
  \BibitemOpen
  \bibfield  {author} {\bibinfo {author} {\bibfnamefont {D.}~\bibnamefont
  {Bru{\ss}}},\ }\bibfield  {title} {\enquote {\bibinfo {title} {Entanglement
  splitting of pure bipartite quantum states},}\ }\href {\doibase
  10.1103/physreva.60.4344} {\bibfield  {journal} {\bibinfo  {journal} {Phys.
  Rev. A}\ }\textbf {\bibinfo {volume} {60}},\ \bibinfo {pages} {4344}
  (\bibinfo {year} {1999})}\BibitemShut {NoStop}%
\end{thebibliography}%
\end{document}